\documentclass[final,3p,times,twocolumn,authoryear]{elsarticle}

\usepackage{amssymb}
\usepackage{graphicx}
\usepackage[colorlinks=true, linkcolor=blue, citecolor=green, urlcolor=cyan]{hyperref}
\journal{arXiV}

\begin{document}

\begin{frontmatter}

\title{Immersive Virtual Reality Assessments of Working Memory and Psychomotor Skills: A Comparison between Immersive and Non-Immersive Assessments}

\author[ed,acg,nta,nta2]{Panagiotis Kourtesis\corref{cor1}}
\author[ed]{Andrea Lizarraga}
\author[ed]{Sarah E. MacPherson}
\address[ed]{Department of Psychology, The University of Edinburgh, Edinburgh, United Kingdom}
\address[acg]{Department of Psychology, The American College of Greece, Athens, Greece}
\address[nta]{Department of Informatics \& Telecommunications, National and Kapodistrian University of Athens, Athens, Greece}
\address[nta2]{Department of Psychology, National and Kapodistrian University of Athens, Athens, Greece}

\cortext[cor1]{Correspondence: pkourtesis@acg.edu}

\begin{abstract}
\textbf{Objective:} Immersive virtual reality (VR) enhances ecological validity and facilitates intuitive and ergonomic hand interactions for performing neuropsychological assessments. However, its comparability to traditional computerized methods remains unclear. This study investigates the convergent validity, user experience, and usability of VR-based versus PC-based assessments of short-term and working memory, and psychomotor skills, while also examining how demographic and IT-related skills influence performance in both modalities.\\[1mm]
\textbf{Methods:} Sixty-six participants performed the Digit Span Task (DST), Corsi Block Task (CBT), and Deary-Liewald Reaction Time Task (DLRTT) in both VR- and PC-based formats. Participants’ experience in using computers and smartphones, and playing videogames, was considered. User experience and system usability of the formats were also evaluated.\\[1mm]
\textbf{Results:} While performance on DST was similar across modalities, PC assessments enabled better performance on CBT and faster reaction times in DLRTT. Moderate-to-strong correlations between VR and PC versions supported convergent validity. Regression analyses revealed that performance on PC versions was influenced by age, computing, and gaming experience, whereas performance on VR versions was largely independent of these factors, except for gaming experience predicting performance on CBT backward recall. Moreover, VR assessments received higher ratings for user experience and usability than PC-based assessments.\\[1mm]
\textbf{Conclusion:} Immersive VR assessments provide an engaging alternative to traditional computerized methods, with minimal reliance on prior IT experience and demographic factors. This resilience to individual differences suggests that VR may offer a more equitable and accessible platform for cognitive assessment. Future research should explore the long-term reliability of VR-based assessments.
\end{abstract}

\begin{keyword}
Virtual Reality \sep Neuropsychological Tests \sep Cognitive Function \sep Working Memory \sep Psychomotor Performance \sep Reaction Time \sep Human-Computer Interaction \sep Usability Testing
\end{keyword}

\end{frontmatter}



\section{Introduction}
Cognitive abilities underpin a wide spectrum of human behaviour, ranging from intricate problem-solving and learning tasks to routine, everyday activities. Within this domain, working memory and psychomotor skills have attracted extensive interest in research and clinical contexts due to their profound impact on functional independence and overall quality of life \citep{Alloway2009,Baddeley1992,Deary2005}. Broadly, working memory enables the temporary storage and manipulation of information, forming the backbone of higher-level processes such as reasoning, language comprehension, and coherent response formation \citep{Gathercole2008,Raghubar2010,Rogers2011}. Psychomotor skills, by contrast, integrate sensory input, cognitive appraisal, and motor output, allowing humans to navigate their environment effectively---whether through basic object manipulation or more complex tasks like driving or operating machinery \citep{Chaiken2000}. Both working memory and psychomotor abilities are susceptible to age-related decline and to neuropathological conditions such as stroke or dementia, reinforcing the need for robust, reliable assessment tools \citep{Alloway2009,Hatem2016}.

Traditionally, neuropsychological evaluations of these constructs have relied on well-established assessment tools and their computerized versions such as the Digit Span Task (DST; \citep{Ramsay1995, Woods2011}), Corsi Block Task (CBT; \citep{Corsi1973,Kessels2000}), and reaction-time measures like the Deary–Liewald Reaction Time Task (DLRTT; \citep{Deary2011}). These assessments offer diagnostic clarity and standardized administration protocols, which is particularly advantageous for identifying cognitive impairment associated with neurological conditions such as Alzheimer’s disease and stroke \citep{Alloway2009,Petersen2004}. The DST evaluates verbal short-term or working memory by requiring individuals to repeat or manipulate sequences of digits \citep{Baddeley1992}, whereas the CBT assesses visuospatial short-term or working memory by calling for participants to replicate block sequences in a forward or backward order \citep{Corsi1973}. Reaction-time tasks, whether simple (SRT) or choice-based (CRT), examine psychomotor coordination and have been linked to fluid intelligence, everyday motor functioning, and even mortality risk \citep{DearyDer2005b,Der2006,Jensen2006}.

Despite their proven utility, the computerized versions of these traditional assessments face significant limitations in capturing the complexity of real-world cognition. They predominantly rely on two-dimensional, highly controlled stimuli and environments that diverge substantially from the multifaceted, three-dimensional challenges people encounter in daily life \citep{KourtesisMacPherson2021,Parsons2015,Rizzo2004}. For instance, tasks like repeating digit sequences or selecting squares on a screen do not reflect the distractor-loaded contexts and spatial characteristics of everyday scenarios. Consequently, these tests may have limited ecological validity, reducing their capacity to predict real-life performance (veridicality) and/or emulate daily experiences (verisimilitude) \citep{Spooner2006,Suchy2024}. While their simplicity and standardization help isolate specific cognitive processes, these instruments cannot replicate the embodied performance of tasks in a three-dimensional, 360-degree, real-world setting, thus overlooking important visuospatial and motor aspects of everyday functioning.

An additional concern relates to digital literacy confounds associated with computerized assessment. A growing body of evidence suggests that individuals with higher familiarity in video gaming, computing, or smartphone use often outperform less tech-savvy participants on computerized tasks \citep{Bauer2012,Feldstein1999,Iverson2009}. Technology-proficient individuals benefit from enhanced fine motor control, faster reaction times, and more adept navigation through on-screen interfaces \citep{Borecki2013}. Consequently, traditional computerized assessments requiring fine motor skills to press keys and/or control a mouse may inadvertently measure digital proficiency rather than purely cognitive constructs. Such a confound is particularly problematic when testing older adults or individuals from lower socioeconomic or educational backgrounds, who may lack consistent exposure to computers or gaming devices \citep{Bauer2012,Feldstein1999}. Thus, while computerized tasks facilitate automated data collection and scoring, they can introduce biases that undermine test fairness and inclusivity.

Immersive Virtual Reality (VR) emerges as a promising solution to these challenges, enabling researchers and clinicians to preserve the structured control of laboratory settings while introducing greater ecological validity. By using head-mounted displays, motion tracking, and optional haptic feedback, VR can simulate complex, three-dimensional environments where participants interact via naturalistic and intuitive embodied interactions instead of keyboards or mice \citep{Kourtesis2020}. This shift may not only lessen the advantage conferred by prior computer experience but also offer a more intuitive means of completing cognitive tasks, as well as potentially improving usability, user experience, and data fidelity \citep{KourtesisMacPherson2021,Zaidi2018}. Immersive VR neuropsychological assessments, such as the Virtual Reality Everyday Assessment Lab \citep{Kourtesis2021} and Nesplora Aquarium \citep{Climent2021}, have demonstrated robust validity. Notably, such immersive formats appear to reduce test fatigue, sustain participant attention, and accurately measure cognitive processes ranging from working memory to executive functions \citep{Kourtesis2020,Rizzo2004}.

Another strength of VR-based methods lies in their capacity to systematically manipulate environmental factors. For example, clinicians can introduce or remove distractors, vary object placement, and alter time pressures in ways that are challenging to replicate in physical or 2D computerized settings \citep{Parsons2015,Rizzo2004}. As VR technology evolves, advanced features like eye-tracking, motion sensors, and multisensory integration (e.g., auditory or haptic cues) promise to reveal even more sophisticated insights into the cognitive and motor abilities involved in task performance \citep{Kim2020,Kourtesis2020,Makinen2022}. This level of detail could help pinpoint whether performance delays stem from slowed visual processing, decreased attention, or impaired motor execution, thereby informing more targeted interventions.

Considering these potential benefits, the present study seeks to directly compare immersive VR-based and traditional computerized versions of three widely used neuropsychological assessments---the DST, CBT, and DLRTT---in terms of convergent validity, performance, user experience, and usability. A further aim is to determine whether gaming or computing experience exerts less influence on VR outcomes, thus evaluating VR’s promise in mitigating technology-based bias. If confirmed, these findings would bolster the case for broader integration of VR in both clinical and research settings, ultimately transforming how practitioners diagnose and treat cognitive impairments across populations with diverse digital literacy levels.

\section{Methods}

\subsection{Participants}
Sixty-six participants (38 women) aged 18 to 45 years (M = 27.89, SD = 4.88), with 12–25 years of education (M = 16.65, SD = 2.70), were recruited through posters, email lists, social media, and participant pools from the University of Edinburgh and the American College of Greece. Ethical approval was obtained from the PPLS Research Ethics Committee of the University of Edinburgh (318-2223/8), ensuring accordance with Helsinki Declaration. All participants provided written informed consent. Each participant received £15 (or the €18 equivalent) as compensation for their time.

\subsection{Materials}

\subsubsection{Hardware and Software}
VR tasks were administered using an HTC Vive Pro Eye headset with built-in eye tracking. This system exceeds minimum recommended hardware specifications for reducing cybersickness \citep{Kourtesis2019}. VR software was developed in Unity 2019.3.f1 \citep{Unity2020}, following ergonomic guidelines \citep{ISO2007} and best practices for VR in neuropsychology \citep{Kourtesis2020}. Interactions were managed via the SteamVR SDK, allowing participants to use hand controllers for naturalistic interaction. A Windows 11 Pro laptop with an Intel Core i9 CPU, 128 GB RAM, and a GeForce GTX 1060 Ti graphics card was used. Audio prompts were generated with Amazon Polly, and spatial audio was facilitated by the SteamAudio plugin. Computerized tasks were hosted on the PsyToolkit platform \citep{Stoet2010,Stoet2017}.

\subsubsection{Questionnaires}
\paragraph{Demographic and IT Skills Questionnaire.} A survey gathered basic demographic information, including age (in years), education (in years of formal schooling), and sex (male/female), and assessed computing and smartphone application experience using two 6‐point Likert-scale items (one on usage frequency, 1 = Never; 6 = Every day, and one on perceived ability, 1 = No Skill; 6 = Expert). The items for each domain were summed to produce a single experience score.

\paragraph{Gaming Skill Questionnaire (GSQ).} The GSQ \citep{Zioga2024, Zioga2024b} was employed to assess participants’ expertise across six gaming genres (i.e., sports games, first-person shooter games, role-playing games, action–adventure games, strategy games, and puzzle games) by asking two 6‐point Likert-scale items per genre: one on frequency of play (1 = Never; 6 = Everyday) and one on self-perceived expertise (1 = No Skill; 6 = Expert). Each genre-specific score was computed by summing these two items, and a total gaming skill score was obtained by summing across all genres.

\paragraph{User Experience Questionnaire (UEQ).} A shortened, 8-item version of the UEQ \citep{Schrepp2017} was used to evaluate subjective impressions of task interfaces (VR vs. PC). Participants rated paired adjectives (e.g., “supportive–obstructive”) on a 7-point Likert scale. The sum of all items yielded a single user experience score with a maximum of 56.

\paragraph{System Usability Scale (SUS).} The SUS \citep{Brooke1995,Brooke2013} is a widely used, 10-item questionnaire measuring perceived usability on a 5-point Likert scale ranging from “Strongly Disagree” to “Strongly Agree.” Sample items include “I found the system unnecessarily complex” and “I felt confident using the system.” Responses are converted into a final 0–50 score, with higher values denoting better usability.

\paragraph{Cybersickness in Virtual Reality Questionnaire (CSQ-VR).} Cybersickness symptoms (e.g., nausea, dizziness, eyestrain) were assessed using the CSQ-VR \citep{Kourtesis2023} on a 7-point scale (1 = No Symptoms, 7 = Intense Symptoms), with a maximum score of 42.

\subsubsection{Cognitive and Psychomotor Tasks}
\begin{figure}[h]
    \centering
    \includegraphics[width=0.45\textwidth]{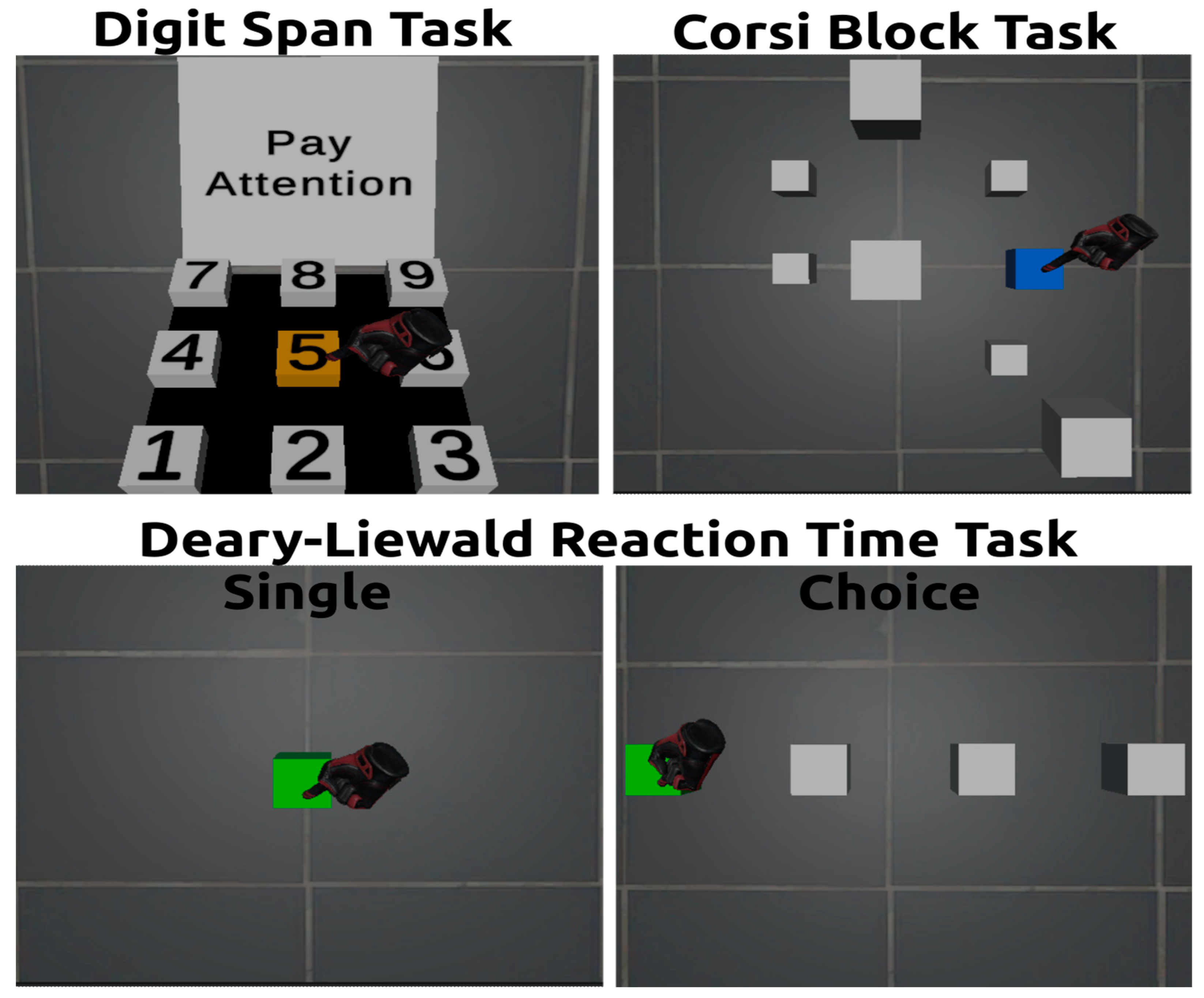}
    \caption{VR Versions of Digit Span Task, Corsi Block Task, and Deary–Liewald Reaction Time Task. Adapted from \citet{Kourtesis2023}.}
    \label{fig:tasks}
\end{figure}

All VR tasks (see Figure~\ref{fig:tasks}) conformed to ISO 9241-400:2007 ergonomic guidelines \citep{ISO2007}, which included adjusting object heights and distances for natural reach. Interactions involved intuitive hand gestures rather than keystrokes, aligning with recommended practices for VR-based cognitive testing \citep{Kourtesis2020}. Participants received neutral audio prompts and visual feedback (green for correct, orange for incorrect). A video overview of the VR versions of the tasks is available at the following links: \href{https://www.youtube.com/watch?v=1H8cqci-lFs}{DST}, \href{https://www.youtube.com/watch?v=MLilvkyMt-g}{CBT}, and \href{https://www.youtube.com/watch?v=wXdrt0PjNsk}{DLRTT} (accessed on January 20, 2025).

\paragraph{Digit Span Task (DST).} The computerized version of the DST \citep{Woods2011,Woods2011b} was adapted based on the original version by Ramsay and Reynolds (1995) and the version included in the Wechsler Adult Intelligence Scale \citep{Wechsler2008}. Participants heard a series of digits at 2-second intervals through headphones and then recalled the sequences in forward or backward order, either by selecting digits using mouse clicks on a computerized keypad or tapping on a virtual keypad. The task began with a two-digit sequence, with the sequence length increasing by one after at least one correct trial; two consecutive errors at the same length ended the task. The total score combined the maximum span achieved and the number of correctly recalled sequences, with a maximum of 20 points for both forward and backward recall.

\paragraph{Corsi Block Task (CBT).} The computerized version of the CBT \citep{Fischer2001} was based on the original wooden version of the Corsi Block Task \citep{Corsi1973} to measure short-term and working memory for spatial sequences. The CBT involves squares/cubes that change color sequentially to depict the order of the sequence. Apart from differences in dimensions (2D vs. 3D) and response methods (mouse clicks vs. virtual touching), the PC and VR versions shared comparable administration and scoring. In VR, 27 cubes were arranged in three-dimensional space, although only 9 appeared in each trial. A subset of these cubes would change colour sequentially, prompting participants to replicate the pattern in either forward or backward order. The task began with a sequence of two cubes, and two sequences at the same length were presented; the sequence length increased by one when at least one sequence was correctly recalled or when the maximum span (7) was reached. The total CBT score was the sum of the highest span achieved and the number of sequences correctly recalled, with a maximum of 20 points for both forward and backward recall \citep{Gould1990}.

\paragraph{Deary–Liewald Reaction Time Task (DLRTT).} The DLRTT \citep{Deary2011} is a computerized measure of psychomotor skills (i.e., eye-hand coordination) that assesses both Simple Reaction Time (SRT) and Choice Reaction Time (CRT). In the VR-SRT, participants touched a cube as quickly as possible when it changed color from white to blue (20 trials). In the VR-CRT, one of four cubes changed color at random (40 trials) and participants were required to touch the changed cube as quickly as possible. The PC version displayed similar stimuli on a monitor and required participants to press the space bar for SRT or specific keyboard keys (e.g., “z”, “x”, “,”, “.”) for CRT. In both versions, mean reaction times (RT) were computed. Notably, the VR version, using integrated eye-tracking, can measure both the time required for users to visually detect a target (attentional time) and the interval between target detection and selection (motor time).

\subsection{Procedure}
All sessions took place in a laboratory at the University of Edinburgh or the American College of Greece. Participants completed a demographic and technological skills questionnaire, as well as the GSQ to indicate their familiarity with video games, smartphones, and computers. Subsequently, each participant performed three cognitive/psychomotor tasks in both VR and PC formats. Half of the participants completed the VR tasks first, while the other half completed the PC tasks first; within each modality, the task order was counterbalanced. Upon completing the VR tasks, participants filled out the CSQ-VR to assess changes in cybersickness. Finally, participants rated both modalities using the UEQ and SUS. Brief breaks were provided between tasks and modalities. The session lasted 60–90 minutes.

\subsection{Statistical Analyses}
All analyses were performed using R (version 4.1.3) \citep{r_core_team_r_2024} and the \texttt{psych} package \citep{Revelle2022} in RStudio \citep{Rstudio2024}. Each outcome variable underwent Shapiro–Wilk tests for normality; variables deviating significantly from normality were transformed using the \texttt{bestNormalize} package \citep{Peterson2020} and then scaled into z-scores for subsequent analyses.

To assess convergent validity, Pearson’s correlations were computed between the VR and PC scores for each task. Paired samples t-tests were conducted with format (VR vs. PC) as a within-subjects factor to examine effects on user experience, system usability, and task performance. Finally, regression analyses were performed to determine whether age, education, computing experience, smartphone experience, and gaming experience predicted performance on the VR and PC tasks. Initially, correlations among these variables were assessed. Next, individual single-predictor models were constructed for each variable and compared to a null model. Only predictors that significantly improved model performance over the null were compared against one another to identify the best-performing variable. Additional predictors were then added incrementally, retaining the best predictor(s) from the previous stage and introducing one new predictor at each step. At every stage, analyses of variance were used to compare models in terms of $R^2$ improvement, while monitoring the variance inflation factor to rule out multicollinearity.

\section{Results}

Participants reported relatively high experience with computing and smartphone applications, with means (SD, range) of 10.21 (1.25, 4--12) and 10.46 (1.24, 6--12), respectively. Videogaming experience showed greater variability, with a mean of 34.6 (12.41, 22--72). Participants reported absent to very mild cybersickness symptoms in VR. Table~\ref{tab:performance} summarizes user experience, system usability, and cognitive task performance across assessment modalities.

\begin{table}[!h]
\caption{User Experience, System Usability, and Performance Across Formats}
\resizebox{\columnwidth}{1.4in}{%
\label{tab:performance}
\begin{tabular}{lcccccc}
\hline
Test & Format & Mean & SD & Minimum & Maximum \\
\hline
UEQ (Max=56)  & PC & 39.27 & 8.91 & 14 & 55 \\
          & VR & 46.12 & 6.79 & 26 & 56 \\
SUS (Max=50)  & PC & 40.38 & 6.54 & 23 & 49 \\
          & VR & 42.64 & 5.47 & 26 & 50 \\
DSF (Max=20)  & PC & 16.38 & 2.83 & 8  & 20 \\
          & VR & 16.09 & 3.35 & 0  & 20 \\
DSB (Max=20)  & PC & 14.85 & 3.46 & 3  & 20 \\
          & VR & 14.83 & 3.33 & 8  & 20 \\
CBF (Max=20)  & PC & 15.79 & 4.39 & 5  & 20 \\
          & VR & 14.85 & 2.64 & 10 & 20 \\
CBB (Max=20)  & PC & 15.52 & 4.64 & 5  & 20 \\
          & VR & 14.82 & 2.16 & 9  & 20 \\
SRT (seconds)   & PC & 0.27  & 0.05 & 0.20 & 0.46 \\
          & VR & 0.48  & 0.10 & 0.27 & 0.69 \\
CRT (seconds)   & PC & 0.42  & 0.07 & 0.30 & 0.60 \\
          & VR & 0.56  & 0.10 & 0.36 & 0.82 \\
CRT -- AT (s) & VR & 0.29  & 0.10 & 0.18 & 0.42 \\
CRT -- MT (s) & VR & 0.23  & 0.11 & 0.06 & 0.46 \\
\hline
\end{tabular}
}
\begin{flushleft}
\footnotesize \textit{Note.}PC = Computerized Assessment, VR = Virtual Reality Assessment, UEQ = User Experience Questionnaire, SUS = System Usability Scale, DSF = Digit Span Task – Forward Recall, DSB = Digit Span Task – Backward Recall, CBF = Corsi Block Task – Forward Recall, CBB = Corsi Block Task – Backward Recall, SRT = Deary Liewald Single Reaction Time Task, CRT= Deary Liewald Choice Reaction Time Task, AT = Attention Time, MT = Motor Time.
\end{flushleft}
\end{table}

\subsection{Convergent Validity between VR and PC Versions}
Across all tasks, the VR-based measures demonstrated positive correlations with their established PC-based counterparts, indicating convergent validity. For the Digit Span Test, VR-based forward recall was strongly correlated with PC-based forward recall, $r(64) = 0.679$, $p < .001$, while the backward recall correlation was moderately strong ($r(64) = 0.516$, $p < .001$). The Corsi Block-Tapping Test showed a small-to-moderate relationship for forward recall ($r(64) = 0.291$, $p = .009$) and a moderate correlation for backward recall ($r(64) = 0.445$, $p < .001$). For the Deary–Liewald Reaction Time Test, the VR-based SRT was modestly associated with the PC-based equivalent ($r(64) = 0.264$, $p = .016$), while the CRT exhibited a robust correlation ($r(64) = 0.591$, $p < .001$).

\subsection{Comparison of Formats: User Experience, System Usability, and Performance}
A paired-samples t-test on the UEQ revealed a significant difference between modalities, $t(65) = -7.01$, $p < .001$, Cohen’s $d = -0.86$, with higher ratings for VR compared to PC. Similarly, SUS scores differed significantly between VR and PC, $t(65) = -5.93$, $p < .001$, $d = -0.73$, again with higher usability ratings for VR. For the performance scores, there were no significant differences between the two modalities for DST forward ($t(65) = 0.63$, $p = .532$, $d = 0.08$) or backward recall ($t(65) = -0.04$, $p = .966$, $d = -0.01$). However, CBT forward and backward recall yielded significantly higher scores on the PC compared to the VR versions [$t(65) = 2.46$, $p = .017$, $d = 0.30$, and $t(65) = 2.65$, $p = .010$, $d = 0.33$, respectively]. Reaction time tasks also showed notable differences favoring PC administration: SRTs, $t(65) = -15.79$, $p < .001$, $d = -1.94$, and CRTs, $t(65) = -12.69$, $p < .001$, $d = -1.56$.

\begin{table}[!h]
\caption{Correlations: Demographics, IT Experience, and Performance on Tests}
\resizebox{\columnwidth}{2in}{%
\fontsize{10pt}{12pt}
\label{tab:correlations}
\begin{tabular}{lcccccc}
\hline
 & & \multicolumn{5}{c}{Experience Variables} \\
Test & Measure & Age & Education & Computing XP & Smartphone XP & Videogaming XP \\
\hline
DSF -- PC & Pearson's $r$ & 0.18 & -0.02 & 0.07 & 0.19 & 0.12 \\
          & $p$-value      & .151  & .877  & .566  & .130  & .341 \\
DSF -- VR & Pearson's $r$ & 0.15 & -0.02 & 0.15 & 0.18 & 0.21 \\
          & $p$-value      & 0.238 & .861  & .074  & .153  & .098 \\
CBF -- PC & Pearson's $r$ & 0.18 & -0.16 & 0.16 & 0.10 & 0.24 \\
          & $p$-value      & 0.151 & .189  & .210  & .425  & \textbf{.050} \\
CBF -- VR & Pearson's $r$ & -0.01& 0.08  & 0.21 & 0.10 & 0.10 \\
          & $p$-value      & .976  & .548  & .097  & .399  & .424 \\
SRT -- PC & Pearson's $r$ & 0.11 & 0.06  & -0.24& -0.04& -0.26 \\
          & $p$-value      & .371  & .608  & \textbf{.049}  & .725  & \textbf{.033} \\
SRT -- VR & Pearson's $r$ & 0.16 & 0.15  & -0.05& -0.08& -0.14 \\
          & $p$-value      & .198  & .219  & .683  & .530  & .252 \\
CRT -- PC & Pearson's $r$ & 0.22 & 0.12  & -0.40& -0.20& -0.36 \\
          & $p$-value      & .071  & .339  & \textbf{$<$.001} & .116 & \textbf{.003} \\
CRT -- VR & Pearson's $r$ & -0.07& 0.04  & -0.14& -0.23& -0.22 \\
          & $p$-value      & .573  & .749  & .261  & .061  & .071 \\
CRT -- AT & Pearson's $r$ & 0.10 & -0.03 & 0.01 & -0.08& -0.15 \\
          & $p$-value      & .432  & .823  & .994  & .550  & .222 \\
CRT -- MT & Pearson's $r$ & 0.14 & 0.07  & -0.02& -0.20& -0.23 \\
          & $p$-value      & .275  & .596  & .878  & .108  & .067 \\
DSB -- PC & Pearson's $r$ & 0.31 & 0.12  & 0.05 & 0.14 & 0.13 \\
          & $p$-value      & \textbf{.013}  & .334  & .677  & .260  & .307 \\
DSB -- VR & Pearson's $r$ & -0.11& 0.01  & 0.05 & 0.11 & -0.01 \\
          & $p$-value      & .385  & .939  & .704  & .363  & .989 \\
CBB -- PC & Pearson's $r$ & 0.11 & -0.08 & 0.29 & 0.30 & 0.39 \\
          & $p$-value      & .366  & .517  & \textbf{.018}  & \textbf{.015}  & \textbf{.001} \\
CBB -- VR & Pearson's $r$ & 0.13 & 0.13  & 0.18 & 0.32 & 0.42 \\
          & $p$-value      & .287  & .282  & .155  & \textbf{.009}  & \textbf{$<$.001} \\
\hline
\end{tabular}
}
\begin{flushleft}
\footnotesize \textit{Note.} Significant correlations are displayed in bold; PC = Computerized Assessment, VR = Virtual Reality Assessment, XP = Experience, DSF = Digit Span Task – Forward Recall, DSB = Digit Span Task – Backward Recall, CBF = Corsi Block Task – Forward Recall, CBB = Corsi Block Task – Backward Recall, SRT = Deary–Liewald Single Reaction Time Task, CRT = Deary–Liewald Choice Reaction Time Task, AT = Attention Time, MT = Motor Time.
\end{flushleft}
\end{table}
\subsection{Predictors of Performance: Demographics and IT Skills}
Correlation analyses (see Table~\ref{tab:correlations}) revealed that participant age, computing experience, and videogaming experience were substantially associated with performance on several PC-based tasks, whereas for VR tasks these factors were generally non-significant—except for smartphone and videogaming experience, which were significantly and positively correlated with performance on the VR version of CBB.

These patterns were further reflected in the linear regression analyses (see Table~\ref{tab:regression}). Age emerged as a significant predictor of performance on the PC versions of DSB and CRT, while computing experience significantly predicted PC CRT. Videogaming experience was the most frequent significant predictor of performance on the PC versions of CBF, CBB, SRT, and CRT, with the best model explaining 27\% of the variance on the PC version of CRT. In contrast, the null models best explained performance on the VR versions of most tests, except that videogaming experience also predicted performance on the VR-based CBB.

\begin{table}[!h]
\caption{Best Linear Regression Models: Demographics \& IT Skills as Predictors of Performance on Cognitive Tests/Tasks}
\resizebox{\columnwidth}{1.4in}{%
\fontsize{10pt}{12pt}
\label{tab:regression}
\begin{tabular}{lcccc}
\hline
Predicted & Predictors & $\beta$ Coefficient & $p$-value & $R^2$ \\
\hline
DSF -- VR   & Null Model         & --   & --     & -- \\
DSF -- PC   & Null Model         & --   & --     & -- \\
DSB -- VR   & Null Model         & --   & --     & -- \\
DSB -- PC   & Age                & 0.30 & .013   & 0.10 \\
CBF -- VR   & Null Model         & --   & --     & -- \\
CBF -- PC   & Videogaming XP     & 0.17 & .033   & 0.07 \\
CBB -- VR   & Videogaming XP     & 0.25 & $<$.001& 0.17 \\
CBB -- PC   & Videogaming XP     & 0.49 & $<$.001& 0.15 \\
SRT -- VR   & Null Model         & --   & --     & -- \\
SRT -- PC   & Videogaming XP     & -0.17& .033   & 0.07 \\
CRT -- VR   & Null Model         & --   & --     & -- \\
CRT -- AT   & Null Model         & --   & --     & -- \\
CRT -- MT   & Null Model         & --   & --     & -- \\
CRT -- PC   & Age                & 0.14 & .018   & 0.27 \\
            & Videogaming XP    & -0.14& .037   & -- \\
            & Computing XP      & -0.15& .026   & -- \\
\hline
\end{tabular}
}
\begin{flushleft}
\footnotesize \textit{Note.} PC = Computerized Assessment, VR = Virtual Reality Assessment, XP = Experience, DSF = Digit Span Task – Forward Recall, DSB = Digit Span Task – Backward Recall, CBF = Corsi Block Task – Forward Recall, CBB = Corsi Block Task – Backward Recall, SRT = Deary–Liewald Single Reaction Time Task, CRT = Deary–Liewald Choice Reaction Time Task, AT = Attention Time, MT = Motor Time.
\end{flushleft}
\end{table}
\section{Discussion}
The present study examined the usability, convergent validity, and performance differences between VR-based and traditional PC-based neuropsychological assessments of working memory and psychomotor skills. Our findings show that VR assessments achieve convergent validity with PC-based tests while offering enhanced engagement with assessments that better resemble the complexity and cognitive demands of the real world. Importantly, regression analyses revealed that VR performance was less influenced by demographic factors and IT-related skills—such as computing and gaming experience—than PC-based tasks. These results suggest that VR can provide a more inclusive, realistic, and less biased platform for neuropsychological evaluation.

\subsection{Validity, User Experience, and Usability of VR Adaptations of Traditional Tasks}
The introduction of 3D spatial components and naturalistic interactions in VR tasks may invoke cognitive processes that more accurately mirror everyday problem-solving, navigation, and decision-making \citep{KourtesisMacPherson2021,Parsons2015}. This could also explain the performance differences between VR and PC versions, where faster reaction times on the DLRTT and better visuospatial memory on the CBT were observed. For example, the VR version of the CBT required encoding and recalling spatial sequences along the $z$-axis, adding depth perception that engages real-world-like navigation processes more than the 2D PC version. Similarly, embodied interactions for target selection, albeit intuitive and ergonomic, require longer movement trajectories compared to key pressing. In line with other validation studies of VR neuropsychological assessments \citep[e.g.,][]{Climent2021,Kourtesis2021}, moderate-to-strong correlations were found between VR and PC performance, adequately supporting the convergent validity of the VR versions. Therefore, while VR introduces extra cognitive and/or motor demands, these do not compromise the accuracy of the cognitive functioning measurements.

Consistent with benefits reported in prior VR neuropsychological assessment studies \citep{KourtesisMacPherson2021,Parsons2015}, our VR assessments received significantly higher user experience and usability ratings than PC methods. This enhanced user experience likely stems from the immersive nature of VR \citep{Makinen2022,Slater2018} as well as its ability to emulate real-world dimensions and challenges \citep{Slater2009,SlaterSanchezVives2016}, thereby better capturing the complexity and cognitive demands of everyday tasks (i.e., verisimilitude) \citep{KourtesisMacPherson2021,Parsons2015}. Furthermore, system usability—as modulated by factors highlighted in recent studies \citep{Kim2020,Makinen2022}—plays a crucial role, especially in clinical populations \citep{Tuena2020} and older adults \citep{Abeele2021,ShaoLee2020}. Given that the VR versions were designed to offer ergonomic \citep{Kourtesis2022} and naturalistic \citep{Kourtesis2020} interactions, the high usability ratings are in line with findings from healthy young and older adult populations as well as clinical groups.

\subsection{Effects of Demographics and IT Skills on Performance}
Regression analyses indicated that demographic factors and IT-related skills, such as age, computing, and gaming experience, significantly influenced PC-based task performance, whereas their impact on VR assessments was minimal. For instance, while videogaming experience modestly predicted performance on the VR version of the CBT backward recall, this relationship was weaker than expected and absent for VR forward recall---in contrast with the PC version where these relationships were significantly stronger. Additionally, although gaming and computing experience accounted for a considerable amount of variance in PC-based reaction time tasks, these factors did not significantly affect performance on VR-based equivalents, suggesting that intuitive VR interactions mitigate advantages conferred by prior digital literacy. These findings are consistent with studies on traditional computerized assessments \citep[e.g.,][]{Feldstein1999,Iverson2009} and those focusing on motor tasks in VR \citep{Kourtesis2022,Zaidi2018}.

The effects of technological proficiency on DLRTT performance further demonstrated how such proficiency can bias test outcomes. In the PC version, both SRT and CRT were strongly predicted by gaming experience, with gaming, computing, and age explaining up to 27\% of the variance on CRT---indicating that traditional computerized tests may conflate cognitive measures with prior technology exposure. In contrast, the VR version showed no significant effect of IT-related experience, suggesting a purer measure of cognitive and motor function. A key advancement in the VR version of the DLRTT was the incorporation of eye-tracking, which allowed for the separation of attentional time (time to detect a visual stimulus) from motor time (time to execute a response). Notably, neither attentional nor motor time was affected by prior IT or gaming experience, enhancing the precision of psychomotor assessments and providing deeper insights into specific delays or inefficiencies. Given that the DLRTT is also employed in cognitive ageing research \citep{Deary2005,Der2006} and can predict age-related disorders and mortality \citep{DearyDer2005b,Der2006}, further examination of the VR version in these contexts is warranted.

\subsection{Implications for Neuropsychological Assessment Methodologies}
The broader implications of our findings extend to neuropsychological testing and research into the effects of videogaming or IT skills on cognitive performance. Specifically, traditional computerized tasks may conflate cognitive assessment with technological proficiency \citep[e.g.,][]{Feldstein1999,Iverson2009}, as also highlighted by the joint position of the American Academy of Clinical Neuropsychology and the National Academy of Neuropsychology \citep{Bauer2012}. Consequently, conventional tests might inadvertently capture differences in fine motor skills rather than pure cognitive abilities. Adopting VR-based testing formats, which minimize such confounds, may therefore provide a more accurate assessment of cognitive constructs. In line with previous work demonstrating that non-gamers can perform comparably to gamers in immersive VR environments \citep{Kourtesis2022,Zaidi2018}, our findings suggest that VR-based tasks require less digital familiarity, offering a purer measure of cognitive function.

By reducing biases related to demographic factors and IT proficiency, VR assessments can foster more inclusive testing environments. This aligns with the criteria of the AACN and NAN concerning cultural, experiential, and disability factors \citep{Bauer2012}. Additionally, the enhanced engagement reported in immersive VR tasks may reduce fatigue and boredom during testing, which is particularly beneficial for populations such as children, older adults, or individuals with cognitive impairments \citep{Borghetti2023,Chiossi2025,GonzalezErena2025}.

\subsection{Limitations and Future Directions}
Despite promising results, our predominantly young and tech-savvy sample limits the generalizability of our findings to older adults or less technologically familiar populations. Future research should involve more diverse samples to assess whether the reduced reliance on IT skills and increased ecological validity observed in VR persist across different user groups. In addition, integrating multisensory inputs (e.g., auditory, haptic) and examining how VR’s spatial and motor demands affect various populations could further refine task design. Longitudinal studies are also needed to evaluate the reliability and sensitivity of VR-based assessments over time, particularly for early detection of cognitive decline or monitoring of neurological conditions. Finally, exploring VR’s utility in clinical populations (e.g., those with neurodegenerative diseases) may pave the way for targeted, adaptive interventions that enhance diagnostic precision and therapeutic outcomes.

\subsection{Conclusion}
This study highlights the potential of immersive VR-based assessments as engaging alternatives to traditional computerized neuropsychological tests. By introducing three-dimensional spatial components, naturalistic interactions, and eye-tracking metrics, VR assessments not only replicate established cognitive measures but also minimize biases associated with prior IT proficiency. As VR technology continues to advance, it may pave the way for the next generation of cognitive testing and intervention tools, offering comprehensive, accurate, and highly engaging platforms that closely mirror the complexities of real-world cognitive function.

\section*{Acknowledgments} 
The VR versions of the Digit Span Task, Corsi Block Task, and Deary–Liewald Reaction Time Task were designed and developed by Dr. Panagiotis Kourtesis. The VR software may freely be shared with third parties for clinical and/or research purposes upon request.

\section*{Financial Support}
The study received no external funding.

\section*{Conflicts of Interest}
The authors declare no conflicts of interest.
\bibliographystyle{elsarticle-harv}
\bibliography{main}

\begin{thebibliography}{61}
\expandafter\ifx\csname natexlab\endcsname\relax\def\natexlab#1{#1}\fi
\providecommand{\url}[1]{\texttt{#1}}
\providecommand{\href}[2]{#2}
\providecommand{\path}[1]{#1}
\providecommand{\DOIprefix}{doi:}
\providecommand{\ArXivprefix}{arXiv:}
\providecommand{\URLprefix}{URL: }
\providecommand{\Pubmedprefix}{pmid:}
\providecommand{\doi}[1]{\href{http://dx.doi.org/#1}{\path{#1}}}
\providecommand{\Pubmed}[1]{\href{pmid:#1}{\path{#1}}}
\providecommand{\bibinfo}[2]{#2}
\ifx\xfnm\relax \def\xfnm[#1]{\unskip,\space#1}\fi
\bibitem[{Abeele et~al.(2021)Abeele, Schraepen, Huygelier, Gillebert, Gerling and Van~Ee}]{Abeele2021}
\bibinfo{author}{Abeele, V.V.}, \bibinfo{author}{Schraepen, B.}, \bibinfo{author}{Huygelier, H.}, \bibinfo{author}{Gillebert, C.}, \bibinfo{author}{Gerling, K.}, \bibinfo{author}{Van~Ee, R.}, \bibinfo{year}{2021}.
\newblock \bibinfo{title}{Immersive virtual reality for older adults: Empirically grounded design guidelines}.
\newblock \bibinfo{journal}{ACM Transactions on Accessible Computing} \bibinfo{volume}{14}.
\newblock \DOIprefix\doi{10.1145/3470743}.
\bibitem[{Alloway et~al.(2009)Alloway, Gathercole, Kirkwood and Elliott}]{Alloway2009}
\bibinfo{author}{Alloway, T.P.}, \bibinfo{author}{Gathercole, S.E.}, \bibinfo{author}{Kirkwood, H.}, \bibinfo{author}{Elliott, J.}, \bibinfo{year}{2009}.
\newblock \bibinfo{title}{The cognitive and behavioral characteristics of children with low working memory}.
\newblock \bibinfo{journal}{Child Development} \bibinfo{volume}{80}, \bibinfo{pages}{606--621}.
\newblock \DOIprefix\doi{10.1111/j.1467-8624.2009.01282.x}.
\bibitem[{Baddeley(1992)}]{Baddeley1992}
\bibinfo{author}{Baddeley, A.}, \bibinfo{year}{1992}.
\newblock \bibinfo{title}{Working memory}.
\newblock \bibinfo{journal}{Science} \bibinfo{volume}{255}, \bibinfo{pages}{556--559}.
\newblock \DOIprefix\doi{10.1126/science.1736359}.
\bibitem[{Bauer et~al.(2012)Bauer, Iverson, Cernich, Binder, Ruff and Naugle}]{Bauer2012}
\bibinfo{author}{Bauer, R.M.}, \bibinfo{author}{Iverson, G.L.}, \bibinfo{author}{Cernich, A.N.}, \bibinfo{author}{Binder, L.M.}, \bibinfo{author}{Ruff, R.M.}, \bibinfo{author}{Naugle, R.I.}, \bibinfo{year}{2012}.
\newblock \bibinfo{title}{Computerized neuropsychological assessment devices: Joint position paper of the american academy of clinical neuropsychology and the national academy of neuropsychology}.
\newblock \bibinfo{journal}{Archives of Clinical Neuropsychology} \bibinfo{volume}{27}, \bibinfo{pages}{362--373}.
\newblock \DOIprefix\doi{10.1093/arclin/acs027}.
\bibitem[{Borecki et~al.(2013)Borecki, Tolstych and Pokorski}]{Borecki2013}
\bibinfo{author}{Borecki, L.}, \bibinfo{author}{Tolstych, K.}, \bibinfo{author}{Pokorski, M.}, \bibinfo{year}{2013}.
\newblock \bibinfo{title}{Computer games and fine motor skills}, in: \bibinfo{booktitle}{Respiratory Regulation - Clinical Advances}, \bibinfo{publisher}{Springer Netherlands}. pp. \bibinfo{pages}{343--348}.
\bibitem[{Borghetti et~al.(2023)Borghetti, Zanobini, Natola, Ottino, Parenti, Brugada-Ramentol, Jalali and Bozorgzadeh}]{Borghetti2023}
\bibinfo{author}{Borghetti, D.}, \bibinfo{author}{Zanobini, C.}, \bibinfo{author}{Natola, I.}, \bibinfo{author}{Ottino, S.}, \bibinfo{author}{Parenti, A.}, \bibinfo{author}{Brugada-Ramentol, V.}, \bibinfo{author}{Jalali, H.}, \bibinfo{author}{Bozorgzadeh, A.}, \bibinfo{year}{2023}.
\newblock \bibinfo{title}{Evaluating cognitive performance using virtual reality gamified exercises}.
\newblock \bibinfo{journal}{Frontiers in Virtual Reality} \bibinfo{volume}{4}.
\newblock \DOIprefix\doi{10.3389/frvir.2023.1153145}.
\bibitem[{Brooke(1995)}]{Brooke1995}
\bibinfo{author}{Brooke, J.}, \bibinfo{year}{1995}.
\newblock \bibinfo{title}{Sus: A quick and dirty usability scale}.
\newblock \bibinfo{journal}{Usability Eval. Ind.} \bibinfo{volume}{189}.
\bibitem[{Brooke(2013)}]{Brooke2013}
\bibinfo{author}{Brooke, J.}, \bibinfo{year}{2013}.
\newblock \bibinfo{title}{Sus: A retrospective}.
\newblock \bibinfo{journal}{Journal of Usability Studies} \bibinfo{volume}{8}, \bibinfo{pages}{29--40}.
\bibitem[{Chaiken et~al.(2000)Chaiken, Kyllonen and Tirre}]{Chaiken2000}
\bibinfo{author}{Chaiken, S.R.}, \bibinfo{author}{Kyllonen, P.C.}, \bibinfo{author}{Tirre, W.C.}, \bibinfo{year}{2000}.
\newblock \bibinfo{title}{Organization and components of psychomotor ability}.
\newblock \bibinfo{journal}{Cognitive Psychology} \bibinfo{volume}{40}, \bibinfo{pages}{198--226}.
\newblock \DOIprefix\doi{10.1006/cogp.1999.0729}.
\bibitem[{Chiossi et~al.(2025)Chiossi, Ou, Gerhardt, Putze and Mayer}]{Chiossi2025}
\bibinfo{author}{Chiossi, F.}, \bibinfo{author}{Ou, C.}, \bibinfo{author}{Gerhardt, C.}, \bibinfo{author}{Putze, F.}, \bibinfo{author}{Mayer, S.}, \bibinfo{year}{2025}.
\newblock \bibinfo{title}{Designing and evaluating an adaptive virtual reality system using eeg frequencies to balance internal and external attention states}.
\newblock \bibinfo{journal}{International Journal of Human-Computer Studies} \bibinfo{volume}{196}, \bibinfo{pages}{103433}.
\newblock \DOIprefix\doi{10.1016/j.ijhcs.2024.103433}.
\bibitem[{Climent et~al.(2021)Climent, Rodríguez, García, Areces, Mejías, Aierbe, Moreno, Cueto, Castellá and Feli~González}]{Climent2021}
\bibinfo{author}{Climent, G.}, \bibinfo{author}{Rodríguez, C.}, \bibinfo{author}{García, T.}, \bibinfo{author}{Areces, D.}, \bibinfo{author}{Mejías, M.}, \bibinfo{author}{Aierbe, A.}, \bibinfo{author}{Moreno, M.}, \bibinfo{author}{Cueto, E.}, \bibinfo{author}{Castellá, J.}, \bibinfo{author}{Feli~González, M.}, \bibinfo{year}{2021}.
\newblock \bibinfo{title}{New virtual reality tool (nesplora aquarium) for assessing attention and working memory in adults: A normative study}.
\newblock \bibinfo{journal}{Applied Neuropsychology: Adult} \bibinfo{volume}{28}, \bibinfo{pages}{403--415}.
\newblock \DOIprefix\doi{10.1080/23279095.2019.1646745}.
\bibitem[{Corsi(1973)}]{Corsi1973}
\bibinfo{author}{Corsi, P.M.}, \bibinfo{year}{1973}.
\newblock \bibinfo{title}{Human memory and the medial temporal region of the brain}.
\newblock Ph.D. thesis. ProQuest Information \& Learning. \bibinfo{address}{US}.
\bibitem[{Deary and Der(2005a)}]{Deary2005}
\bibinfo{author}{Deary, I.J.}, \bibinfo{author}{Der, G.}, \bibinfo{year}{2005}a.
\newblock \bibinfo{title}{Reaction time, age, and cognitive ability: Longitudinal findings from age 16 to 63 years in representative population samples}.
\newblock \bibinfo{journal}{Aging, Neuropsychology, and Cognition} \bibinfo{volume}{12}, \bibinfo{pages}{187--215}.
\newblock \DOIprefix\doi{10.1080/13825580590969235}.
\bibitem[{Deary and Der(2005b)}]{DearyDer2005b}
\bibinfo{author}{Deary, I.J.}, \bibinfo{author}{Der, G.}, \bibinfo{year}{2005}b.
\newblock \bibinfo{title}{Reaction time explains iq's association with death}.
\newblock \bibinfo{journal}{Psychological Science} \bibinfo{volume}{16}, \bibinfo{pages}{64--69}.
\newblock \DOIprefix\doi{10.1111/j.0956-7976.2005.00781.x}.
\bibitem[{Deary et~al.(2011)Deary, Liewald and Nissan}]{Deary2011}
\bibinfo{author}{Deary, I.J.}, \bibinfo{author}{Liewald, D.}, \bibinfo{author}{Nissan, J.}, \bibinfo{year}{2011}.
\newblock \bibinfo{title}{A free, easy-to-use, computer-based simple and four-choice reaction time programme: The deary-liewald reaction time task}.
\newblock \bibinfo{journal}{Behavior Research Methods} \bibinfo{volume}{43}, \bibinfo{pages}{258--268}.
\newblock \DOIprefix\doi{10.3758/s13428-010-0024-1}.
\bibitem[{Der and Deary(2006)}]{Der2006}
\bibinfo{author}{Der, G.}, \bibinfo{author}{Deary, I.J.}, \bibinfo{year}{2006}.
\newblock \bibinfo{title}{Age and sex differences in reaction time in adulthood: Results from the united kingdom health and lifestyle survey}.
\newblock \bibinfo{journal}{Psychology and Aging} \bibinfo{volume}{21}, \bibinfo{pages}{62--73}.
\newblock \DOIprefix\doi{10.1037/0882-7974.21.1.62}.
\bibitem[{Feldstein et~al.(1999)Feldstein, Keller, Portman, Durham, Klebe and Davis}]{Feldstein1999}
\bibinfo{author}{Feldstein, S.N.}, \bibinfo{author}{Keller, F.R.}, \bibinfo{author}{Portman, R.E.}, \bibinfo{author}{Durham, R.L.}, \bibinfo{author}{Klebe, K.J.}, \bibinfo{author}{Davis, H.P.}, \bibinfo{year}{1999}.
\newblock \bibinfo{title}{A comparison of computerized and standard versions of the wisconsin card sorting test}.
\newblock \bibinfo{journal}{The Clinical Neuropsychologist} \bibinfo{volume}{13}, \bibinfo{pages}{303--313}.
\newblock \DOIprefix\doi{10.1076/clin.13.3.303.1744}.
\bibitem[{Fischer(2001)}]{Fischer2001}
\bibinfo{author}{Fischer, M.H.}, \bibinfo{year}{2001}.
\newblock \bibinfo{title}{Probing spatial working memory with the corsi blocks task}.
\newblock \bibinfo{journal}{Brain and Cognition} \bibinfo{volume}{45}, \bibinfo{pages}{143--154}.
\newblock \DOIprefix\doi{10.1006/brcg.2000.1221}.
\bibitem[{Gathercole et~al.(2008)Gathercole, Alloway, Kirkwood, Elliott, Holmes and Hilton}]{Gathercole2008}
\bibinfo{author}{Gathercole, S.E.}, \bibinfo{author}{Alloway, T.P.}, \bibinfo{author}{Kirkwood, H.J.}, \bibinfo{author}{Elliott, J.G.}, \bibinfo{author}{Holmes, J.}, \bibinfo{author}{Hilton, K.A.}, \bibinfo{year}{2008}.
\newblock \bibinfo{title}{Attentional and executive function behaviours in children with poor working memory}.
\newblock \bibinfo{journal}{Learning and Individual Differences} \bibinfo{volume}{18}, \bibinfo{pages}{214--223}.
\newblock \DOIprefix\doi{10.1016/j.lindif.2007.10.003}.
\bibitem[{González-Erena et~al.(2025)González-Erena, Fernández-Guinea and Kourtesis}]{GonzalezErena2025}
\bibinfo{author}{González-Erena, P.V.}, \bibinfo{author}{Fernández-Guinea, S.}, \bibinfo{author}{Kourtesis, P.}, \bibinfo{year}{2025}.
\newblock \bibinfo{title}{Cognitive assessment and training in extended reality: Multimodal systems, clinical utility, and current challenges}.
\newblock \bibinfo{journal}{Encyclopedia} \bibinfo{volume}{5}.
\newblock \DOIprefix\doi{10.3390/encyclopedia5010008}.
\bibitem[{Gould and Glencross(1990)}]{Gould1990}
\bibinfo{author}{Gould, J.H.}, \bibinfo{author}{Glencross, D.J.}, \bibinfo{year}{1990}.
\newblock \bibinfo{title}{Do children with a specific reading disability have a general serial-ordering deficit?}
\newblock \bibinfo{journal}{Neuropsychologia} \bibinfo{volume}{28}, \bibinfo{pages}{271--278}.
\newblock \DOIprefix\doi{10.1016/0028-3932(90)90020-O}.
\bibitem[{Hatem et~al.(2016)Hatem, Saussez, Della~Faille, Prist, Zhang, Dispa and Bleyenheuft}]{Hatem2016}
\bibinfo{author}{Hatem, S.M.}, \bibinfo{author}{Saussez, G.}, \bibinfo{author}{Della~Faille, M.}, \bibinfo{author}{Prist, V.}, \bibinfo{author}{Zhang, X.}, \bibinfo{author}{Dispa, D.}, \bibinfo{author}{Bleyenheuft, Y.}, \bibinfo{year}{2016}.
\newblock \bibinfo{title}{Rehabilitation of motor function after stroke: A multiple systematic review focused on techniques to stimulate upper extremity recovery}.
\newblock \bibinfo{journal}{Frontiers in Human Neuroscience} \bibinfo{volume}{10}, \bibinfo{pages}{442}.
\newblock \DOIprefix\doi{10.3389/fnhum.2016.00442}.
\bibitem[{{International Organization for Standardization}(2007)}]{ISO2007}
\bibinfo{author}{{International Organization for Standardization}}, \bibinfo{year}{2007}.
\newblock \bibinfo{title}{ISO 9241-400:2007: Ergonomics of human-system interaction -- Part 400: Principles and requirements for physical input devices}.
\newblock \bibinfo{type}{Technical Report}. {International Organization for Standardization}. \bibinfo{address}{Geneva}.
\bibitem[{Iverson et~al.(2009)Iverson, Brooks, Ashton, Johnson and Gualtieri}]{Iverson2009}
\bibinfo{author}{Iverson, G.L.}, \bibinfo{author}{Brooks, B.L.}, \bibinfo{author}{Ashton, V.L.}, \bibinfo{author}{Johnson, L.G.}, \bibinfo{author}{Gualtieri, C.T.}, \bibinfo{year}{2009}.
\newblock \bibinfo{title}{Does familiarity with computers affect computerized neuropsychological test performance?}
\newblock \bibinfo{journal}{Journal of Clinical and Experimental Neuropsychology} \bibinfo{volume}{31}, \bibinfo{pages}{594--604}.
\newblock \DOIprefix\doi{10.1080/13803390802372125}.
\bibitem[{Jensen(2006)}]{Jensen2006}
\bibinfo{author}{Jensen, A.R.}, \bibinfo{year}{2006}.
\newblock \bibinfo{title}{Clocking the mind: Mental chronometer individual differences}.
\newblock \bibinfo{publisher}{Elsevier}, \bibinfo{address}{Amsterdam, Netherlands}.
\bibitem[{Kessels et~al.(2000)Kessels, Van~Zandvoort, Postma, Kappelle and De~Haan}]{Kessels2000}
\bibinfo{author}{Kessels, R.P.C.}, \bibinfo{author}{Van~Zandvoort, M.J.E.}, \bibinfo{author}{Postma, A.}, \bibinfo{author}{Kappelle, L.J.}, \bibinfo{author}{De~Haan, E.H.F.}, \bibinfo{year}{2000}.
\newblock \bibinfo{title}{The corsi block-tapping task: Standardization and normative data}.
\newblock \bibinfo{journal}{Applied Neuropsychology} \bibinfo{volume}{7}, \bibinfo{pages}{252--258}.
\newblock \DOIprefix\doi{10.1207/S15324826AN0704\_8}.
\bibitem[{Kim et~al.(2020)Kim, Rhiu and Yun}]{Kim2020}
\bibinfo{author}{Kim, Y.M.}, \bibinfo{author}{Rhiu, I.}, \bibinfo{author}{Yun, M.H.}, \bibinfo{year}{2020}.
\newblock \bibinfo{title}{A systematic review of a virtual reality system from the perspective of user experience}.
\newblock \bibinfo{journal}{International Journal of Human–Computer Interaction} \bibinfo{volume}{36}, \bibinfo{pages}{893--910}.
\newblock \DOIprefix\doi{10.1080/10447318.2019.1699746}.
\bibitem[{Kourtesis et~al.(2021)Kourtesis, Collina, Doumas and MacPherson}]{Kourtesis2021}
\bibinfo{author}{Kourtesis, P.}, \bibinfo{author}{Collina, S.}, \bibinfo{author}{Doumas, L.A.}, \bibinfo{author}{MacPherson, S.E.}, \bibinfo{year}{2021}.
\newblock \bibinfo{title}{Validation of the virtual reality everyday assessment lab (vr-eal): An immersive virtual reality neuropsychological battery with enhanced ecological validity}.
\newblock \bibinfo{journal}{Journal of the International Neuropsychological Society} \bibinfo{volume}{27}, \bibinfo{pages}{181--196}.
\newblock \DOIprefix\doi{10.1017/S1355617720000764}.
\bibitem[{Kourtesis et~al.(2019)Kourtesis, Collina, Doumas and MacPherson}]{Kourtesis2019}
\bibinfo{author}{Kourtesis, P.}, \bibinfo{author}{Collina, S.}, \bibinfo{author}{Doumas, L.A.A.}, \bibinfo{author}{MacPherson, S.E.}, \bibinfo{year}{2019}.
\newblock \bibinfo{title}{Technological competence is a pre-condition for effective implementation of virtual reality head mounted displays in human neuroscience: A technological review and meta-analysis}.
\newblock \bibinfo{journal}{Frontiers in Human Neuroscience} \bibinfo{volume}{13}.
\newblock \DOIprefix\doi{10.3389/fnhum.2019.00342}.
\bibitem[{Kourtesis et~al.(2020)Kourtesis, Korre, Collina, Doumas and MacPherson}]{Kourtesis2020}
\bibinfo{author}{Kourtesis, P.}, \bibinfo{author}{Korre, D.}, \bibinfo{author}{Collina, S.}, \bibinfo{author}{Doumas, L.A.A.}, \bibinfo{author}{MacPherson, S.E.}, \bibinfo{year}{2020}.
\newblock \bibinfo{title}{Guidelines for the development of immersive virtual reality software for cognitive neuroscience and neuropsychology: The development of virtual reality everyday assessment lab (vr-eal)}.
\newblock \bibinfo{journal}{Frontiers in Computer Science} \bibinfo{volume}{1}.
\newblock \DOIprefix\doi{10.3389/fcomp.2019.00012}.
\bibitem[{Kourtesis et~al.(2023)Kourtesis, Linnell, Amir, Argelaguet and MacPherson}]{Kourtesis2023}
\bibinfo{author}{Kourtesis, P.}, \bibinfo{author}{Linnell, J.}, \bibinfo{author}{Amir, R.}, \bibinfo{author}{Argelaguet, F.}, \bibinfo{author}{MacPherson, S.E.}, \bibinfo{year}{2023}.
\newblock \bibinfo{title}{Cybersickness in virtual reality questionnaire (csq-vr): A validation and comparison against ssq and vrsq}.
\newblock \bibinfo{journal}{Virtual Worlds} \bibinfo{volume}{2}, \bibinfo{pages}{16--35}.
\newblock \DOIprefix\doi{10.3390/virtualworlds2010002}.
\bibitem[{Kourtesis and MacPherson(2021)}]{KourtesisMacPherson2021}
\bibinfo{author}{Kourtesis, P.}, \bibinfo{author}{MacPherson, S.E.}, \bibinfo{year}{2021}.
\newblock \bibinfo{title}{How immersive virtual reality methods may meet the criteria of the national academy of neuropsychology and american academy of clinical neuropsychology: A software review of the virtual reality everyday assessment lab (vr-eal)}.
\newblock \bibinfo{journal}{Computers in Human Behavior Reports} \bibinfo{volume}{4}, \bibinfo{pages}{100151}.
\bibitem[{Kourtesis et~al.(2022)Kourtesis, Vizcay, Marchal, Pacchierotti and Argelaguet}]{Kourtesis2022}
\bibinfo{author}{Kourtesis, P.}, \bibinfo{author}{Vizcay, S.}, \bibinfo{author}{Marchal, M.}, \bibinfo{author}{Pacchierotti, C.}, \bibinfo{author}{Argelaguet, F.}, \bibinfo{year}{2022}.
\newblock \bibinfo{title}{Action-specific perception \& performance on a fitts's law task in virtual reality: The role of haptic feedback}.
\newblock \bibinfo{journal}{IEEE Transactions on Visualization and Computer Graphics} \bibinfo{volume}{28}, \bibinfo{pages}{3715--3726}.
\newblock \DOIprefix\doi{10.1109/TVCG.2022.3203003}.
\bibitem[{Mäkinen et~al.(2022)Mäkinen, Haavisto, Havola and Koivisto}]{Makinen2022}
\bibinfo{author}{Mäkinen, H.}, \bibinfo{author}{Haavisto, E.}, \bibinfo{author}{Havola, S.}, \bibinfo{author}{Koivisto, J.M.}, \bibinfo{year}{2022}.
\newblock \bibinfo{title}{User experiences of virtual reality technologies for healthcare in learning: An integrative review}.
\newblock \bibinfo{journal}{Behaviour \& Information Technology} \bibinfo{volume}{41}, \bibinfo{pages}{1--17}.
\newblock \DOIprefix\doi{10.1080/0144929X.2020.1788162}.
\bibitem[{Parsons(2015)}]{Parsons2015}
\bibinfo{author}{Parsons, T.D.}, \bibinfo{year}{2015}.
\newblock \bibinfo{title}{Virtual reality for enhanced ecological validity and experimental control in the clinical, affective and social neurosciences}.
\newblock \bibinfo{journal}{Frontiers in Human Neuroscience} \bibinfo{volume}{9}.
\newblock \DOIprefix\doi{10.3389/fnhum.2015.00660}.
\bibitem[{Petersen(2004)}]{Petersen2004}
\bibinfo{author}{Petersen, R.C.}, \bibinfo{year}{2004}.
\newblock \bibinfo{title}{Mild cognitive impairment as a diagnostic entity}.
\newblock \bibinfo{journal}{Journal of Internal Medicine} \bibinfo{volume}{256}, \bibinfo{pages}{183--194}.
\newblock \DOIprefix\doi{10.1111/j.1365-2796.2004.01388.x}.
\bibitem[{Peterson and Cavanaugh(2020)}]{Peterson2020}
\bibinfo{author}{Peterson, R.A.}, \bibinfo{author}{Cavanaugh, J.E.}, \bibinfo{year}{2020}.
\newblock \bibinfo{title}{Ordered quantile normalization: A semiparametric transformation built for the cross-validation era}.
\newblock \bibinfo{journal}{Journal of Applied Statistics} \bibinfo{volume}{47}, \bibinfo{pages}{2312--2327}.
\newblock \DOIprefix\doi{10.1080/02664763.2019.1630372}.
\bibitem[{{Posit team}(2024)}]{Rstudio2024}
\bibinfo{author}{{Posit team}}, \bibinfo{year}{2024}.
\newblock \bibinfo{title}{RStudio: Integrated Development Environment for R}.
\newblock \bibinfo{publisher}{Posit Software, PBC}, \bibinfo{address}{Boston, MA}.
\bibitem[{{R Core Team}(2024)}]{r_core_team_r_2024}
\bibinfo{author}{{R Core Team}}, \bibinfo{year}{2024}.
\newblock \bibinfo{title}{R: A language and environment for statistical computing}.
\bibitem[{Raghubar et~al.(2010)Raghubar, Barnes and Hecht}]{Raghubar2010}
\bibinfo{author}{Raghubar, K.P.}, \bibinfo{author}{Barnes, M.A.}, \bibinfo{author}{Hecht, S.A.}, \bibinfo{year}{2010}.
\newblock \bibinfo{title}{Working memory and mathematics: A review of developmental, individual difference, and cognitive approaches}.
\newblock \bibinfo{journal}{Learning and Individual Differences} \bibinfo{volume}{20}, \bibinfo{pages}{110--122}.
\newblock \DOIprefix\doi{10.1016/j.lindif.2009.10.005}.
\bibitem[{Ramsay and Reynolds(1995)}]{Ramsay1995}
\bibinfo{author}{Ramsay, M.C.}, \bibinfo{author}{Reynolds, C.R.}, \bibinfo{year}{1995}.
\newblock \bibinfo{title}{Separate digits tests: A brief history, a literature review, and a reemination of the factor structure of the test of memory and learning (tomal)}.
\newblock \bibinfo{journal}{Neuropsychology Review} \bibinfo{volume}{5}, \bibinfo{pages}{151--171}.
\newblock \DOIprefix\doi{10.1007/BF02214760}.
\bibitem[{Revelle(2022)}]{Revelle2022}
\bibinfo{author}{Revelle, W.}, \bibinfo{year}{2022}.
\newblock \bibinfo{title}{psych: {Procedures} for {Psychological}, {Psychometric}, and {Personality} {Research}}.
\newblock \bibinfo{publisher}{Northwestern University}, \bibinfo{address}{Evanston, Illinois}.
\newblock \URLprefix \url{https://CRAN.R-project.org/package=psych}.
\bibitem[{Rizzo et~al.(2004)Rizzo, Schultheis, Kerns and Mateer}]{Rizzo2004}
\bibinfo{author}{Rizzo, A.A.}, \bibinfo{author}{Schultheis, M.}, \bibinfo{author}{Kerns, K.A.}, \bibinfo{author}{Mateer, C.}, \bibinfo{year}{2004}.
\newblock \bibinfo{title}{Analysis of assets for virtual reality applications in neuropsychology}.
\newblock \bibinfo{journal}{Neuropsychological Rehabilitation} \bibinfo{volume}{14}, \bibinfo{pages}{207--239}.
\newblock \DOIprefix\doi{10.1080/09602010343000183}.
\bibitem[{Rogers et~al.(2011)Rogers, Hwang, Toplak, Weiss and Tannock}]{Rogers2011}
\bibinfo{author}{Rogers, M.}, \bibinfo{author}{Hwang, H.}, \bibinfo{author}{Toplak, M.}, \bibinfo{author}{Weiss, M.}, \bibinfo{author}{Tannock, R.}, \bibinfo{year}{2011}.
\newblock \bibinfo{title}{Inattention, working memory, and academic achievement in adolescents referred for attention deficit/hyperactivity disorder (adhd)}.
\newblock \bibinfo{journal}{Child Neuropsychology} \bibinfo{volume}{17}, \bibinfo{pages}{444--458}.
\newblock \DOIprefix\doi{10.1080/09297049.2010.544648}.
\bibitem[{Schrepp et~al.(2017)Schrepp, Hinderks and Thomaschewski}]{Schrepp2017}
\bibinfo{author}{Schrepp, M.}, \bibinfo{author}{Hinderks, A.}, \bibinfo{author}{Thomaschewski, J.}, \bibinfo{year}{2017}.
\newblock \bibinfo{title}{Construction of a benchmark for the user experience questionnaire (ueq)}.
\newblock \bibinfo{journal}{International Journal of Interactive Multimedia and Artificial Intelligence} \bibinfo{volume}{4}, \bibinfo{pages}{40--44}.
\newblock \DOIprefix\doi{10.25968/opus-3397}.
\bibitem[{Shao and Lee(2020)}]{ShaoLee2020}
\bibinfo{author}{Shao, D.}, \bibinfo{author}{Lee, I.J.}, \bibinfo{year}{2020}.
\newblock \bibinfo{title}{Acceptance and influencing factors of social virtual reality in the urban elderly}.
\newblock \bibinfo{journal}{Sustainability (Switzerland)} \bibinfo{volume}{12}, \bibinfo{pages}{1--19}.
\newblock \DOIprefix\doi{10.3390/su12229345}. \bibinfo{note}{number: 22}.
\bibitem[{Slater(2009)}]{Slater2009}
\bibinfo{author}{Slater, M.}, \bibinfo{year}{2009}.
\newblock \bibinfo{title}{Place illusion and plausibility can lead to realistic behaviour in immersive virtual environments}.
\newblock \bibinfo{journal}{Philosophical Transactions of the Royal Society B: Biological Sciences} \bibinfo{volume}{364}, \bibinfo{pages}{3549--3557}.
\newblock \DOIprefix\doi{10.1098/rstb.2009.0138}. \bibinfo{note}{number: 1535}.
\bibitem[{Slater(2018)}]{Slater2018}
\bibinfo{author}{Slater, M.}, \bibinfo{year}{2018}.
\newblock \bibinfo{title}{Immersion and the illusion of presence in virtual reality}.
\newblock \bibinfo{journal}{British Journal of Psychology} \bibinfo{volume}{109}, \bibinfo{pages}{431--433}.
\newblock \URLprefix \url{https://doi.org/10.1111/bjop.12305}, \DOIprefix\doi{10.1111/bjop.12305}. \bibinfo{note}{number: 3 Publisher: John Wiley \& Sons, Ltd}.
\bibitem[{Slater and Sanchez-Vives(2016)}]{SlaterSanchezVives2016}
\bibinfo{author}{Slater, M.}, \bibinfo{author}{Sanchez-Vives, M.V.}, \bibinfo{year}{2016}.
\newblock \bibinfo{title}{Enhancing {Our} {Lives} with {Immersive} {Virtual} {Reality}}.
\newblock \bibinfo{journal}{Frontiers in Robotics and AI} \bibinfo{volume}{3}.
\bibitem[{Spooner and Pachana(2006)}]{Spooner2006}
\bibinfo{author}{Spooner, D.M.}, \bibinfo{author}{Pachana, N.A.}, \bibinfo{year}{2006}.
\newblock \bibinfo{title}{Ecological validity in neuropsychological assessment: A case for greater consideration in research with neurologically intact populations}.
\newblock \bibinfo{journal}{Archives of Clinical Neuropsychology} \bibinfo{volume}{21}, \bibinfo{pages}{327--337}.
\newblock \DOIprefix\doi{10.1016/j.acn.2006.04.004}.
\bibitem[{Stoet(2010)}]{Stoet2010}
\bibinfo{author}{Stoet, G.}, \bibinfo{year}{2010}.
\newblock \bibinfo{title}{Psytoolkit: A software package for programming psychological experiments using linux}.
\newblock \bibinfo{journal}{Behavior Research Methods} \bibinfo{volume}{42}, \bibinfo{pages}{1096--1104}.
\newblock \DOIprefix\doi{10.3758/BRM.42.4.1096}.
\bibitem[{Stoet(2017)}]{Stoet2017}
\bibinfo{author}{Stoet, G.}, \bibinfo{year}{2017}.
\newblock \bibinfo{title}{Psytoolkit: A novel web-based method for running online questionnaires and reaction-time experiments}.
\newblock \bibinfo{journal}{Teaching of Psychology} \bibinfo{volume}{44}, \bibinfo{pages}{24--31}.
\newblock \DOIprefix\doi{10.1177/0098628316677643}.
\bibitem[{Suchy et~al.(2024)Suchy, DesRuisseaux, Mora, Brothers and Niermeyer}]{Suchy2024}
\bibinfo{author}{Suchy, Y.}, \bibinfo{author}{DesRuisseaux, L.A.}, \bibinfo{author}{Mora, M.G.}, \bibinfo{author}{Brothers, S.L.}, \bibinfo{author}{Niermeyer, M.A.}, \bibinfo{year}{2024}.
\newblock \bibinfo{title}{Conceptualization of the term “ecological validity” in neuropsychological research on executive function assessment: A systematic review and call to action}.
\newblock \bibinfo{journal}{Journal of the International Neuropsychological Society} \bibinfo{volume}{30}, \bibinfo{pages}{499--522}.
\newblock \DOIprefix\doi{10.1017/S1355617723000735}.
\bibitem[{Tuena et~al.(2020)Tuena, Pedroli, Trimarchi, Gallucci, Chiappini, Goulene, Gaggioli, Riva, Lattanzio, Giunco and Stramba-Badiale}]{Tuena2020}
\bibinfo{author}{Tuena, C.}, \bibinfo{author}{Pedroli, E.}, \bibinfo{author}{Trimarchi, P.D.}, \bibinfo{author}{Gallucci, A.}, \bibinfo{author}{Chiappini, M.}, \bibinfo{author}{Goulene, K.}, \bibinfo{author}{Gaggioli, A.}, \bibinfo{author}{Riva, G.}, \bibinfo{author}{Lattanzio, F.}, \bibinfo{author}{Giunco, F.}, \bibinfo{author}{Stramba-Badiale, M.}, \bibinfo{year}{2020}.
\newblock \bibinfo{title}{Usability issues of clinical and research applications of virtual reality in older people: A systematic review}.
\newblock \bibinfo{journal}{Frontiers in Human Neuroscience} \bibinfo{volume}{14}.
\bibitem[{{Unity Technologies}(2020)}]{Unity2020}
\bibinfo{author}{{Unity Technologies}}, \bibinfo{year}{2020}.
\newblock \bibinfo{title}{Unity (version 2019.3.0f1) [computer software]}.
\bibitem[{Wechsler(2008)}]{Wechsler2008}
\bibinfo{author}{Wechsler, D.}, \bibinfo{year}{2008}.
\newblock \bibinfo{title}{Wechsler adult intelligence scale, fourth edition (wais-iv): Technical manual}.
\bibitem[{Woods et~al.(2011a)Woods, Herron, Yund, Hink, Kishiyama and Reed}]{Woods2011b}
\bibinfo{author}{Woods, D.L.}, \bibinfo{author}{Herron, T.J.}, \bibinfo{author}{Yund, E.W.}, \bibinfo{author}{Hink, R.F.}, \bibinfo{author}{Kishiyama, M.M.}, \bibinfo{author}{Reed, B.}, \bibinfo{year}{2011}a.
\newblock \bibinfo{title}{Computerized analysis of error patterns in digit span recall}.
\newblock \bibinfo{journal}{Journal of Clinical and Experimental Neuropsychology} \bibinfo{volume}{33}, \bibinfo{pages}{721--734}.
\newblock \DOIprefix\doi{10.1080/13803395.2010.550602}.
\bibitem[{Woods et~al.(2011b)Woods, Kishiyama, Yund, Herron, Edwards, Poliva, Hink and Reed}]{Woods2011}
\bibinfo{author}{Woods, D.L.}, \bibinfo{author}{Kishiyama, M.M.}, \bibinfo{author}{Yund, E.W.}, \bibinfo{author}{Herron, T.J.}, \bibinfo{author}{Edwards, B.}, \bibinfo{author}{Poliva, O.}, \bibinfo{author}{Hink, R.F.}, \bibinfo{author}{Reed, B.}, \bibinfo{year}{2011}b.
\newblock \bibinfo{title}{Improving digit span assessment of short-term verbal memory}.
\newblock \bibinfo{journal}{Journal of Clinical and Experimental Neuropsychology} \bibinfo{volume}{33}, \bibinfo{pages}{101--111}.
\newblock \DOIprefix\doi{10.1080/13803395.2010.493149}.
\bibitem[{Zaidi et~al.(2018)Zaidi, Duthie, Carr and Maksoud}]{Zaidi2018}
\bibinfo{author}{Zaidi, S.F.M.}, \bibinfo{author}{Duthie, C.}, \bibinfo{author}{Carr, E.}, \bibinfo{author}{Maksoud, S.H.A.E.}, \bibinfo{year}{2018}.
\newblock \bibinfo{title}{Conceptual framework for the usability evaluation of gamified virtual reality environment for non-gamers}, in: \bibinfo{booktitle}{Proceedings of the 16th ACM SIGGRAPH International Conference on Virtual-Reality Continuum and Its Applications in Industry}, \bibinfo{publisher}{Association for Computing Machinery}, \bibinfo{address}{New York, NY, USA}. pp. \bibinfo{pages}{343--348}.
\newblock \DOIprefix\doi{10.1145/3284398.3284431}.
\bibitem[{Zioga et~al.(2024a)Zioga, Ferentinos, Konsolaki, Nega and Kourtesis}]{Zioga2024b}
\bibinfo{author}{Zioga, T.}, \bibinfo{author}{Ferentinos, A.}, \bibinfo{author}{Konsolaki, E.}, \bibinfo{author}{Nega, C.}, \bibinfo{author}{Kourtesis, P.}, \bibinfo{year}{2024}a.
\newblock \bibinfo{title}{Video {Game} {Skills} across {Diverse} {Genres} and {Cognitive} {Functioning} in {Early} {Adulthood}: {Verbal} and {Visuospatial} {Short}-{Term} and {Working} {Memory}, {Hand}–{Eye} {Coordination}, and {Empathy}}.
\newblock \bibinfo{journal}{Behavioral Sciences} \bibinfo{volume}{14}.
\newblock \DOIprefix\doi{10.3390/bs14100874}. \bibinfo{note}{number: 10}.
\bibitem[{Zioga et~al.(2024b)Zioga, Nega, Roussos and Kourtesis}]{Zioga2024}
\bibinfo{author}{Zioga, T.}, \bibinfo{author}{Nega, C.}, \bibinfo{author}{Roussos, P.}, \bibinfo{author}{Kourtesis, P.}, \bibinfo{year}{2024}b.
\newblock \bibinfo{title}{Validation of the {Gaming} {Skills} {Questionnaire} in {Adolescence}: {Effects} of {Gaming} {Skills} on {Cognitive} and {Affective} {Functioning}}.
\newblock \bibinfo{journal}{European Journal of Investigation in Health, Psychology and Education} \bibinfo{volume}{14}, \bibinfo{pages}{722--752}.
\newblock \DOIprefix\doi{10.3390/ejihpe14030048}. \bibinfo{note}{number: 3}.

\end{thebibliography}
\end{document}